# A Generalizable Theory for the Reynolds Stress, Based on the Lagrangian Turbulence Transport

T.-W. Lee

*Mechanical and Aerospace Engineering, SEMTE, Arizona State University, Tempe, AZ, 85287*

**Abstract-** Using the Lagrangian transport of momentum, the Reynolds shear stress can be expressed in terms of basic turbulence parameters. In this view, the Reynolds shear stress represents the lateral transport of streamwise momentum, balanced by the $u'^2$ transport, pressure and shear force terms in the momentum equation. We extend this formalism to other turbulence parameters such as diagonal components of the Reynolds stress tensor, and also to more complex flows (boundary layer flows with adverse pressure gradient and pipe flows with swirl). Data from direct numerical simulations (DNS) are used to validate this full set of turbulent transport equations, exhibiting a good degree of consistency and agreement for all of the components and across different geometries. An example of the use of this full set of turbulence transport equations is shown for jet flows.

T.-W. Lee
Mechanical and Aerospace Engineering, SEMTE
Arizona State University
Tempe, AZ 85287-6106
Email: attwl@asu.edu

**INTRODUCTION**

Determination of the Reynolds stress in terms of root turbulence parameters has profound implications in fluid physics and engineering, as its shear stress component adds to the mean momentum structure. For some simple geometries such as fully-developed pipe flow, prescription of the Reynolds shear stress results directly in the mean velocity profile.

$$\mu \frac{d^2 U}{dy^2} = \frac{dP}{dx} + \rho \frac{d(u'v')}{dy} \qquad (1)$$

Thus, it may be stated that determination of the Reynolds stress is central to solving the turbulence problem. Since many scientific and engineering phenomena involve turbulent flows, different ways of expressing the Reynolds stress in terms of known (or computable) parameters have been devised so that artificial closure is achieved, with varying degrees of success (or inadequacy, depending on one's perspective). The procedure is typically to expand the Reynolds stress or the "turbulence viscosity", in terms of higher-order terms, e.g. turbulence kinetic energy, dissipation, and third-order correlations in the case of Reynolds stress models. In turn, these higher-order terms need to be algebraically modelled as reviewed in some articles (Launder et al., 1976; Mansour et al., 1988; Hanjalic, 1994; Pope, 2011; Hamba, 2013; Inagaki et al., 2019). The models are quite diverse and at times complex, and we defer the detailed discussions to these references since the existing methods have little or no bearing on the current approach.

Recently, we have developed a new formulation for the Reynolds stress based on the Galilean-transformed Navier-Stokes equation (Lee, 2018; Lee, 2020; Lee and Park, 2017), where a simple, explicit expression is derived for the gradient of the Reynolds shear stress, as shown below:

$$\frac{d(u'v')}{dy} = -C_1 U \left[\frac{d(u'^2)}{dy} + \frac{1}{\rho}\frac{d|P|}{dy}\right] + v\frac{d^2 u'}{dy^2} \qquad (2)$$

The derivation can be found in our previous work (e.g. Lee, 2020), and an expanded version below. Eq. 2 is an expression arising from the momentum balance for a control volume moving at the local mean velocity, where the Reynolds shear stress represents the lateral (y-direction) transport of u' momentum (Lee, 2020). This lateral transport is balanced by the longitudinal transport and force terms on the right-hand side (RHS) of Eq. 2. The d/dx gradients are transformed to d/dy, following the displacement of the fluid (Lee, 2018; Lee, 2020). Viscous shear stress due to gradient of the turbulent fluctuation velocity (u' = u'$_{rms}$) is a new concept; however, we can visualize that if there exists a gradient in the mean fluctuation velocity then it will lead to shear stress in the mean in this moving coordinate frame. The shear stress from the mean velocity is considered to be negligible in comparison to this "turbulent shear stress". Once we have the gradient from Eq. 2, then the Reynolds stress itself can be obtained through numerical or algebraic integration. For example, for rectangular channel flows we can integrate Eq. 1 by parts:

$$<u'v'> = -C_1 \left[U u'^2 - \int_0^y u'^2 \frac{dU}{dy} dy\right] + v \frac{du'_{rms}}{dy} \qquad (3)$$

This approach has proven to work quite well, insofar as prescribing the Reynolds shear stress, for channel, boundary layer and jet flows (Lee, 2018; Lee, 2020). We can see that in order to solve for u'v' we introduce yet another turbulence parameter, $u'^2$. A similar transport equation needs to be written for this variable in order to simultaneously solve for $u'^2$, u'v' and U (Lee, 2020).

The current formulation is based on momentum conservation principle in a non-stationary coordinate frame, so that the resulting expression for the Reynolds stress (Eq. 2) contains, and also reveals, the transport processes leading to the turbulence shear stress. For this reason, it should be possible to generalize this approach to other more complex geometries. In this work, we show that this approach can indeed be applied to other geometries such as pipe flow with swirl and boundary-layer flows with adverse pressure gradients, and the turbulence momentum balance checked with DNS (direct numerical simulation) data. Moreover, we extend the Lagrangian transport formalism to other turbulence variables, $u'^2$ and $v'^2$, so that the complete Reynolds stress tensor (diagonal and off-diagonal terms) can be determined. This extension produces a full set of transport equations for the Reynolds stress tensor, so that relatively simple numerical integration can be performed in conjunction with the Reynolds-averaged Navier-Stokes equation (RANS, e.g. Eq. 1) to solve for the turbulent flow structure. Comparisons are made with DNS and experimental data to validate this method, along with the unique hypotheses embodied therein.

**LAGRANGIAN TURBULENCE TRANSPORT**

We can visualize a control volume moving at the local mean velocity, U and V, so that the effects of turbulence fluctuation components (the Reynolds stress) are isolated, as depicted in Fig. 1. Eq. 2 is the transport equation for u' which has already been described (Lee, 2018; Lee, 2020), so we consider two new transport equations for v' and $u'^2$. If the current formalism did not lead to the correct transport balance for these components, then something would be amiss and in any event we need these additional variables since they appear as unknowns in Eq. 2 ($P \sim \rho v^2$).

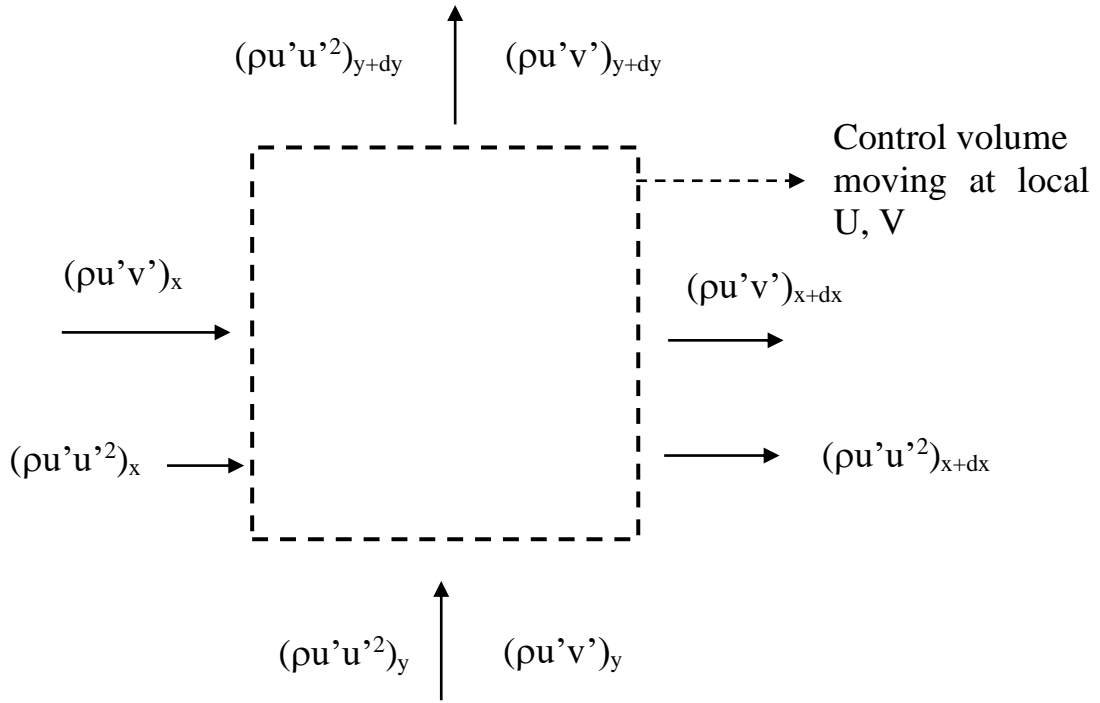

(a)

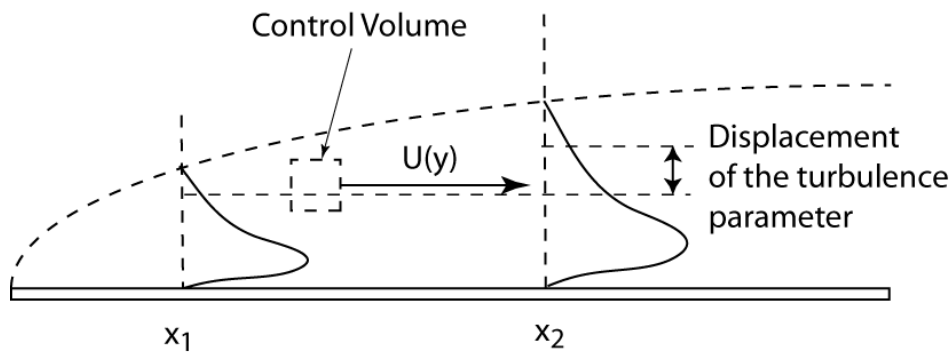

(b)

**Fig. 1. Schematics of the dynamics contained in the Lagrangian turbulence transport: (a) y-momentum (v') and longitudinal kinetic energy balance (u'$^2$) following a control volume, moving at the local mean velocity; and (b) the displacement concept leading to d/dx → d/dy transform.**

From Fig. 1, the transport equations for v' (lateral turbulence momentum) and u'$^2$ (longitudinal turbulence kinetic energy) follow the same logic as the mean momentum equations since U→ 0 and V→0 in this coordinate frame.

Conservation of v' momentum:

$$\frac{d(u'v')}{dx} + \frac{d(v'^2)}{dy} = -\frac{1}{\rho}\frac{dP}{dy} + \nu \frac{d^2 v_{rms}}{dy^2} \quad (4)$$

Conservation of u'$^2$:

$$\frac{d(u'^2 u')}{dx} + \frac{d(u'^2 v')}{dy} = \frac{d(Pu')}{dx} + 2\nu \left(\frac{du'}{dy}\right)^2 \quad (5)$$

Now, we make the d/dx to d/dy transform, which was used to arrive at Eq. 2. This is based on the displacement concept as shown in Fig. 1(b) (Lee, 2018; Lee, 2020). We write this as the first of the several unique hypotheses embodied in this formalism.

Hypothesis 1: $\frac{d}{dx} \rightarrow \pm CU \frac{d}{dy}$ (6),

where C is a constant with the order of magnitude, ~ 1/U$_{ref}$. The +/- sign depends on the flow geometry, i.e. the direction of displacement relative to the reference point. There are no displacement effects for channel flows since the flow is bounded by the walls on both sides. However, we still obtain the above conversion in the spatial gradients using the "probe transform" (Lee, 2019). A brief algebra of this transform is shown in the Appendix. Implicit in this hypothesis is that the both the vectors (u', v') and scalars (pressure, u'$^2$) are transformed in the same manner.

The u²-transport equation (Eq. 5) contains the dreaded triple correlation(s). For Lagrangian transport, an expedient hypothesis is used for the triple correlations, and validated later with DNS data.

Hypothesis 2: $u'^2 v' \approx u' \cdot \overline{u'v'}, \quad \overline{u'v'^2} \approx v' \cdot \overline{u'v'}, etc.$ (7)

This simplistic and expedient product rule decomposes the triple product in terms of existing variables, and leads to unexpected results for u'² profile, as discussed in the next section. The second of the triple product, u'v'², appears in Eq. 5 when the pressure is written as below.

Hypothesis 3: $P \approx -\rho v'^2$ (8)

For channel flows, this relation is exact (Pope, 2012), and we apply it as an approximation for boundary layer flows with and without adverse pressure gradient. Again, the efficacy of these hypotheses is demonstrated with comparison with data in the next section. This pressure expression leads to Pu' → u'v'² in Eq. 5, which is converted to v'(u'v') via hypothesis 2 (Eq. 7). This at first appears to be an over-simplification, but produces to some unexpectedly accurate results. The last hypothesis is that the pressure work done on the moving control volume is mainly due to the mean pressure-u' combination.

Hypothesis 4: $Pressure\ work \approx \dfrac{d(Pu')}{dx}$ (9)

These are evidently new and unique concepts, nonetheless we proceed to write the following set of turbulence transport equations using the above hypotheses. We already showed the transport equation for d(u'v')/dy in Eq. 2, and we just re-write after organizing the constants in $C_{ij}$ format.

$$\frac{d(u'v')}{dy} = -C_{11} U \frac{d(u'^2)}{dy} + C_{12} U \frac{dv'^2}{dy} + C_{13} \frac{d^2 u'}{dy^2} \qquad (10)$$

For v' momentum transport, we have

$$\frac{d(v'^2)}{dy} = -C_{21} U \frac{d(u'v')}{dy} + C_{22} U \frac{dv'^2}{dy} + C_{23} \frac{d^2 v_{rms}}{dy^2} \qquad (11)$$

The second term on the RHS, which is derived from dP/dy in Eq. 4, requires an explanation. As the control volume moves through the boundary layer, it will be subject to additional transport again through the displacement effect (Fig. 1b), similar to $C_{11}$ term in Eq. 10. Inclusion of this term is critical for the correct momentum balance in the current Lagrangian analysis, as shown later.

Finally, for the u'² transport,

$$\frac{d(u'^3)}{dy} = -C_{31} \frac{1}{U} \frac{d(u'v' \cdot u')}{dy} + C_{32} \frac{1}{U} \frac{d(v' \cdot u'v')}{dy} + C_{33} \frac{1}{U} \left(\frac{du'}{dy}\right)^2 \qquad (12)$$

C_{i1} is the displacement constant, used in hypothesis 1. We expect there will be some Reynolds-number dependence for this and other constants, but defer the detailed analysis of the constants to a later study. For now, we optimize the constants based on comparison with data. $C_{i2}$ modifies the pressure term while absorbing the effect of density, and $C_{i3}$ are the kinematic viscosity, $\nu$ ($C_{13}$ and $C_{23}$) or $2\nu$ ($C_{33}$). In the v' transport equation (Eq. 11a), the pressure gradient term is still modified by the mean velocity, attributable to the lateral shift in the pressure field for the control volume. For the u'$^2$ transport (Eq. 12), U appears in the denominator for all the terms on the RHS, since the displacement (or probe) transform was applied to the u'$^2$ transport term on the LHS.

We can apply the same logic to slightly more complex flows than previously considered. For boundary-layer flows with adverse pressure gradients, Eq. 10 can be used, without having to add or adjust any of the terms. The pressure effect is intrinsically manifest through P ~ -ρv'$^2$ (hypothesis 3), i.e., $C_{12}$ term. For axi-symmetric flows containing swirl, we can similarly write the following turbulence momentum transport equations after properly vectorizing the gradients and centripetal pressure in cylindrical coordinates. The logic is similar, and we write the gradient of the main Reynolds shear stress, u$_r$'u$_z$'.

$$\frac{d}{dr}(ru'_r u'_z) = -A_{11} r U_{trans} \frac{d(u'^2_z)}{dr} - A_{12} r U_{trans} \frac{d(u'_\theta u'_z)}{dr} - A_{13} r U_{trans} \frac{dP}{dr} + A_{14} r \frac{d}{dr}\left(r \frac{du'_{z,rms}}{dr}\right)$$

(13)

Here, the centripetal pressure is given by

$$\frac{dP}{dr} = -\rho \frac{U_\theta^2}{r} \qquad (14)$$

The translational velocity in swirl flow is

$$U_{trans} = \sqrt{U_z^2 + U_\theta^2} \qquad (15)$$

For $u_r'u_\theta'$, we have,

$$\frac{d}{dr}(ru_r'u_\theta') = -A_{21}rU_{trans}\frac{d(u_\theta'^2)}{dr} - A_{22}rU_{trans}\frac{d(u_\theta'u_z')}{dr} - A_{33}rU_{trans}\frac{dP}{dr} + A_{34}r\frac{d}{dr}\left(r\frac{du_{\theta,rms}'}{dr}\right)$$

$$(16)$$

Thus, the current formalism allows for extension to other geometries, and for derivation of the transport equations for all the requisite turbulence variables. We demonstrate validity and applicability of this method using available DNS data.

**RESULTS AND DISCUSSION**

**A Full Set of Turbulent Transport Equations**

Let us examine the above full set of turbulence transport equations (Eqs. 10-12) for rectangular channel flows, for which high-fidelity DNS data are readily available, e.g. Graham et al. (2016). We can start with the Reynolds shear stress gradient, $du^+v^+/dy$, as shown in Fig. 2(a). The superscript "+" indicates normalization by the friction velocity, as presented in Graham et al. (2016). The Reynolds stress budgets from Eqs. 10-12 are checked by inputting the DNS data on the RHS and computing the sum of all the terms from appropriated differentials. Then, the left-hand side (LHS) is evaluated separately from the DNS, and compared with the sum from the RHS. When this is done in Eq. 10, we can see that there is a very close agreement between the current theory and DNS data. As discussed in our earlier work (Lee, 2020), it is the triad of momentum terms, longitudinal transport, pressure and viscous forces, that leads the Reynolds shear stress. Thus, intuitively and theoretically, the Reynolds shear stress is determined by the fluctuating momentum components, particularly since $P = -\rho v^2$ in channel flows. We can integrate the gradient of the Reynolds shear stress thus obtained from Eq. 10, to compare directly with $u^+v^+$, in Fig. 2(b). Although there is a good agreement for the gradient between theory and DNS in Fig. 2(a), any minute deviation in the slope can cause appreciable departure during numerical integration. Any error in the gradient is thus magnified, and causes trajectory deviations sending the integrated quantity in the false direction. This would result in accumulated discrepancy from the correct $u^+v^+$. There are two remedies to reduce this kind of numerical integration error: one is to enforce the boundary condition at the other end (e.g. centerline) algebraically, while the second is to integrate from both ends (wall and the centerline) until the solution intersects. This is merely a numerical error correction scheme and stipulation of the boundary conditions, and does not deduct from the theoretical integrity. For the Reynolds shear stress, we opt for the latter method since the small gradient near the centerline is benign to the numerical integration, and can be used for a large section of the channel flow. This "back" integration (inbound from the centerline) is plotted as a dashed line in Fig. 2(b), while the "forward" integration (outbound from the wall) is plotted as a solid line. Comparison with DNS data is quite good, using the above dual integration scheme. Due to the large negative gradient involved near the wall, the forward integration starts to deviate at y/H of only 0.12 and back-integrated solution is used aft of that point.

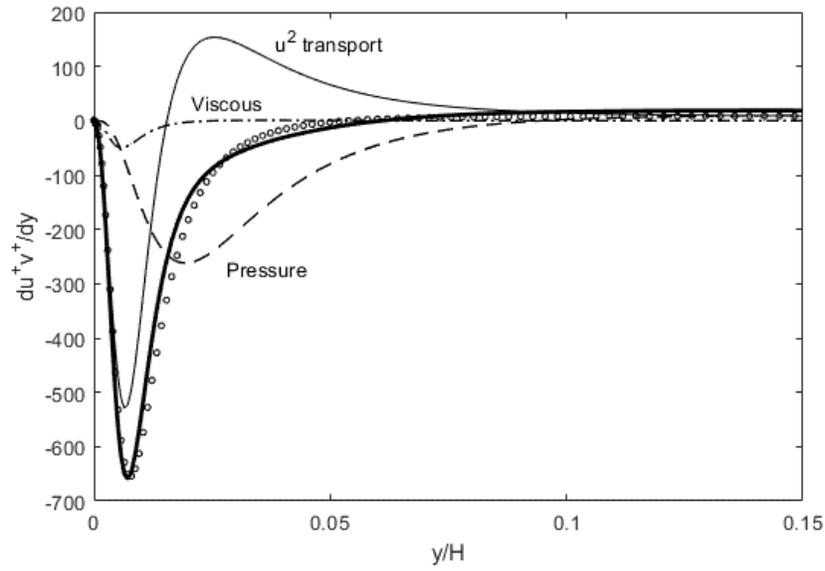

(a)

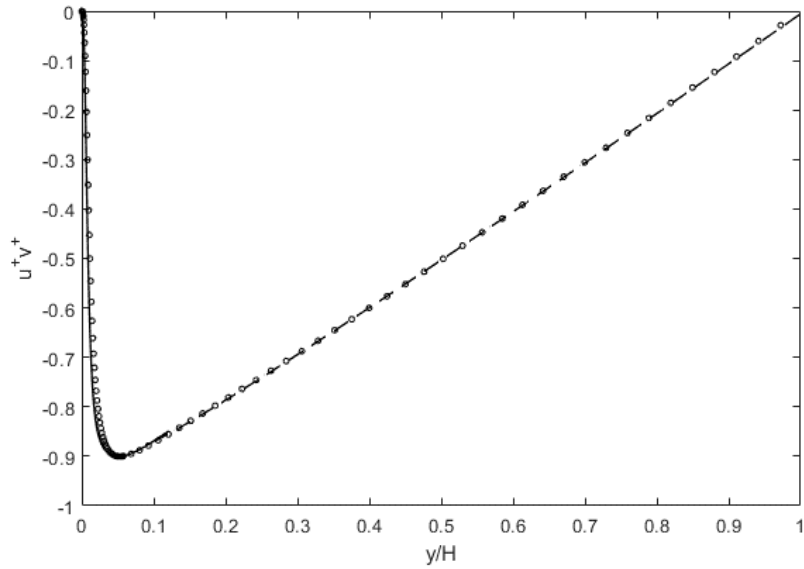

(b)

**Fig. 2. (a) Reynolds shear stress gradient budget (a) for channel flows, consisting of $u'^2$ transport, viscous and pressure terms in Eq. 10; and (b) Comparison of $u^+v^+$, obtained from numerical integration of Eq. 10, with DNS data (Graham et al., 2016).**

For the lateral fluctuating momentum flux, $v^{+2}$, we plot its gradient in Fig. 3(a). Although the gradient obtained using Eq. 11 is structurally correct, the accuracy is not as good as the previous example for $du^+v^+/dy$. We can see that the gradient for $v^{+2}$ is much smaller in magnitude than $du^+v^+/dy$, thus relative numerical accuracy is reduced during computation of the gradient terms. Also, the slope does not quite reach zero at the centerline. Nonetheless, we can still integrate and again compare directly with DNS, in Fig. 3(b). The centerline boundary condition is enforced, this time by using a correction function of the form, $(v^{+2})_{corr} = (v^{+2})_{integrated} + c_v(y/H)$. A linear function of y/H is added to integrated $v^{+2}$, where $c_v = \Delta v^{+2}$ = deviation of the $(v^{+2})_{integrated}$ from $v^{+2}$ at y/H=1. This will ensure that centerline $v^{+2}$ is reached, or close in this case, at y/H=1. Then, the agreement is acceptable, as shown in Fig. 3(b).

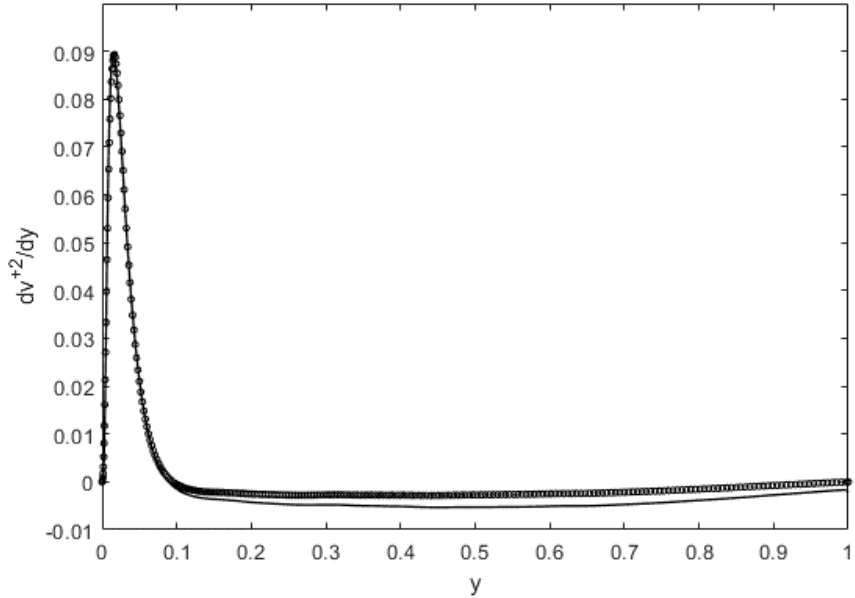

(a)

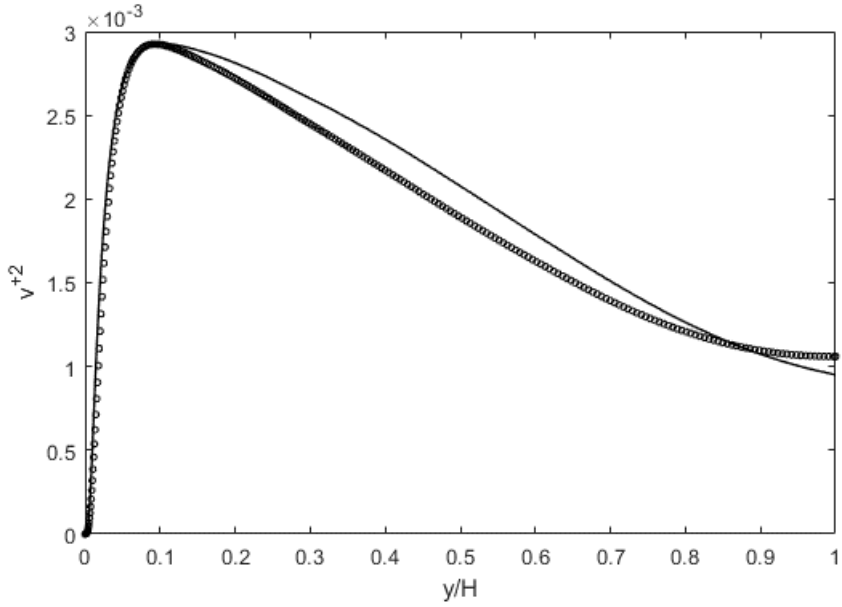

(b)

**Fig. 3.** (a) $dv^{+2}/dy$ budget for channel flows from Eq. 11; and (b) comparison of $v^{+2}$, obtained from numerical integration of Eq. 11, with DNS data (Graham et al., 2016).

For the gradient of the longitudinal kinetic energy flux, $du^{+3}/dy$, the agreement between the theory (Eq. 12) and DNS-derived result is almost exact in Fig. 4(a), somewhat unexpected due to the approximation, $u^{+2}v^{+}=u^{+}(u^{+}v^{+})$, from Eq. 7. This prompts further scrutiny. We zoom in on the near-wall region (y/H= 0 ~ 0.1), and also vary the parameters in Eq. 12, $C_{31}$ and $C_{32}$, relative to what resulted in the theoretical line in Fig. 4(a). Re-calculated gradients with these parametric variations are shown in Fig. 4(b), along with the original theoretical (solid) line, which exhibits very close tracking of the DNS gradient even upon close inspection. Variations of the constants, $C_{31}$ and $C_{32}$, alter the $u^{+3}$ gradient, where $C_{31}$ determines the near-wall peak while $C_{32}$ causes the variation in the secondary "dip" in the gradient. This brief exercise confirms the $u^{+2}$ balance expressed by Eq. 12, and also reveals the dynamics involved in "turbulence production", including the $u^{+2}v^{+}=u^{+}(u^{+}v^{+})$ hypothesis. The near-wall peak evidently means that the x-component of the turbulent kinetic energy, u$^{+2}$, increases quite sharply there. The sensitivity of this peak to C$_{31}$ indicates that that term, C$_{31}$du$^{+2}$v$^+$/dy, is mainly responsible for this surge. Recalling that this is the transport of u$^{+2}$ due to vertical fluctuation, v$^+$, it is the lateral "migration" of u$^{+2}$ that results in a large accumulation of turbulence kinetic energy close to the wall. The secondary dip, then, is due to the pressure work term, suggesting that the built-up kinetic energy is expended by the pressure work, thus distributing it toward the centerline. In summary, the Lagrangian transport analysis succinctly and accurately presents the underlying dynamics of Reynolds stress and turbulent kinetic energy, while confirming the seemingly simple approximation of u$^{+2}$v$^+$=u$^+$(u$^+$v$^+$). A brief analysis for justification of this approximation has been made in Lee (2020), for jet flows.

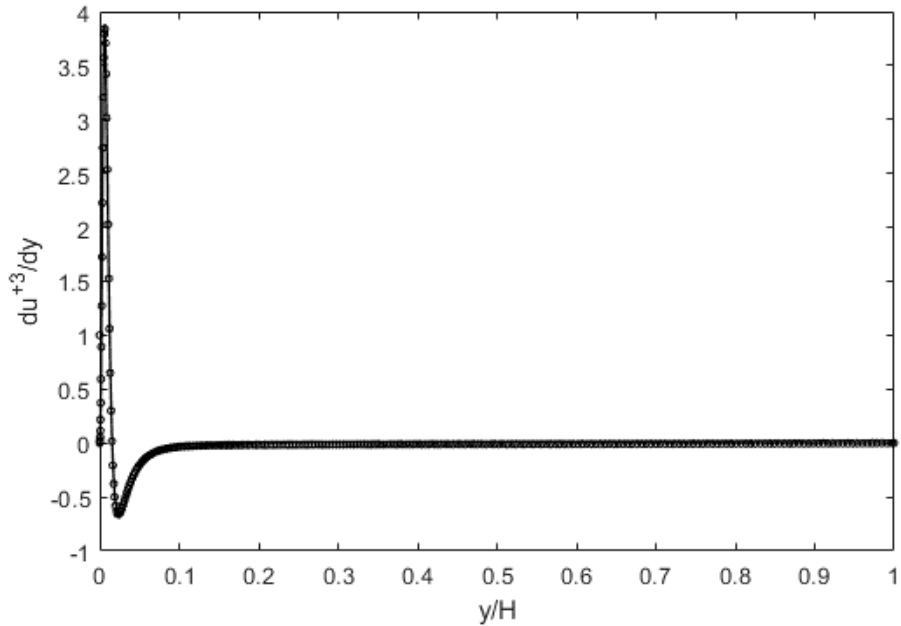

(a)

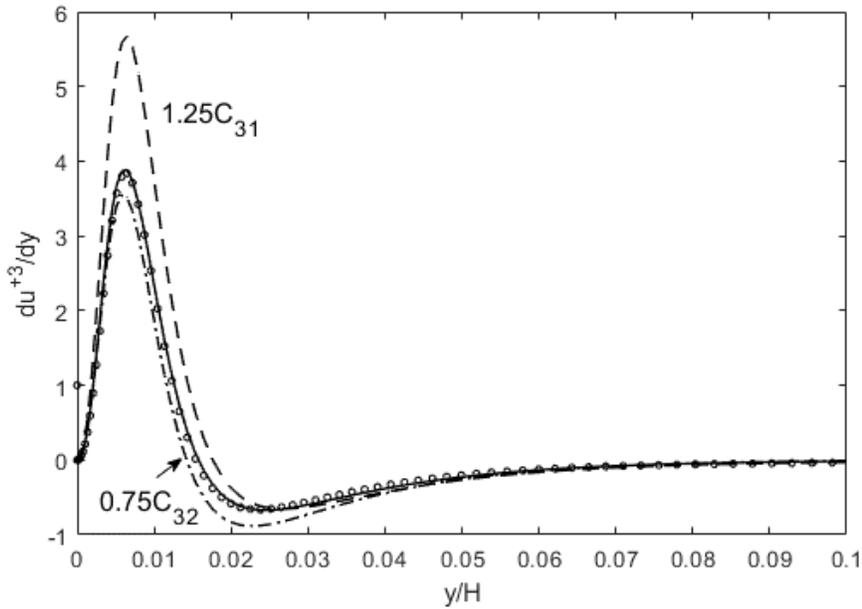

(b)

**Fig. 4. (a) $du^{+3}/dy$ budget for channel flows from Eq. 11; and (b) parametric variation zoomed in close to the wall. DNS data is from Graham et al. (2016).**

The above gradient can again be integrated to yield the spatial distribution of $u^{+2}$, by first finding $u^{+3}$ then setting $u^{+2} = (u^{+3})^{2/3}$. The result is shown in Fig. 5. In spite of the close agreement in $du^{+3}/dy$, the integration is not so forgiving of any minute errors and results in deviation from the DNS data due to sharp peak and large gradient turns in the near-wall region. Moreover, these errors accumulate to cause departure from the centerline (y/H=1) boundary condition, when integrated starting from the wall (y/H=0). To correct for the numerical errors, we again enforce the centerline boundary condition, without any loss of integrity, by setting $(u^{+2})_{corr} = (u^{+2})_{integrated} - c_u(y/H)$. The constant $c_u$ is again $\Delta u^{+2}$ =deviation of the $(u^{+2})_{integrated}$ from $u^{+2}$ at y/H=1. The result is shown below, where due to the above difficulty, there is some + and - deviations from the DNS data in the mid-flow region (y/H=0.1 ~ 1). Nonetheless, the result is not bad considering the difficulty of tracking the sharp peak and gradient changes in the $u'^2$ profiles, which tend to grow in severity with increasing Reynolds number in wall-bounded turbulent flows (Lee, 2019).

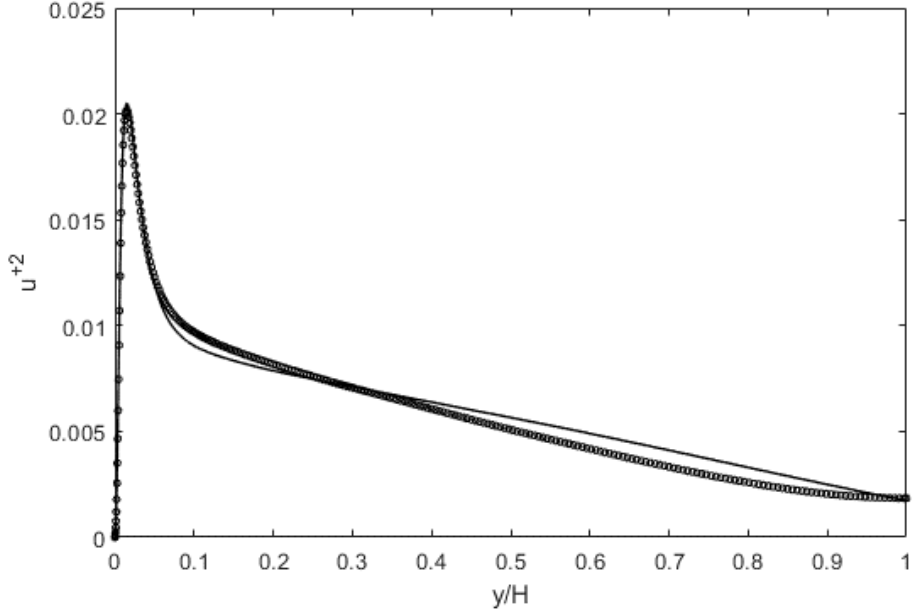

**Fig. 5.** Comparison of $u^{+2}$ profile obtained from numerical integration of Eq. 11, with DNS data (Graham et al., 2016). Boundary conditions are applied both at the wall (y/H=0) and the centerline (y/H=1).

**Applications in Other Flow Geometries**

Now we compare Eq. 10 with the DNS data for adverse-pressure gradient boundary layer flows by Kitsios et al. (2016), for $Re_{\delta 2} = 3500$, as shown in Fig. 6. Figs. 6(a) and (b) are the entire profile across the boundary layer and zoomed-in version for the near-wall region, respectively. We can see that the overall agreement between Eq. 10 and DNS results is good across the boundary layer; however, the undulations away from the wall is mimicked but not exactly followed by Eq. 10. This is due to the inaccurate tracing of the DNS data, leading to some errors when taking the gradient of the transcribed data. This kind of numerical differentiation errors is also evident in Figure 6(b) as well, where a close inspection of the near-wall region again shows that Eq. 1 tracks the DNS data reasonably well, albeit with some small spikes and undulations. At the inflection point ($y^+\sim40$), there is some deviation from DNS data. Nonetheless, Figs. 6(a) and (b) provide some solid grounds for Eq. 10, where the adverse pressure gradient effect is manifest through the $C_{12}$ pressure term.

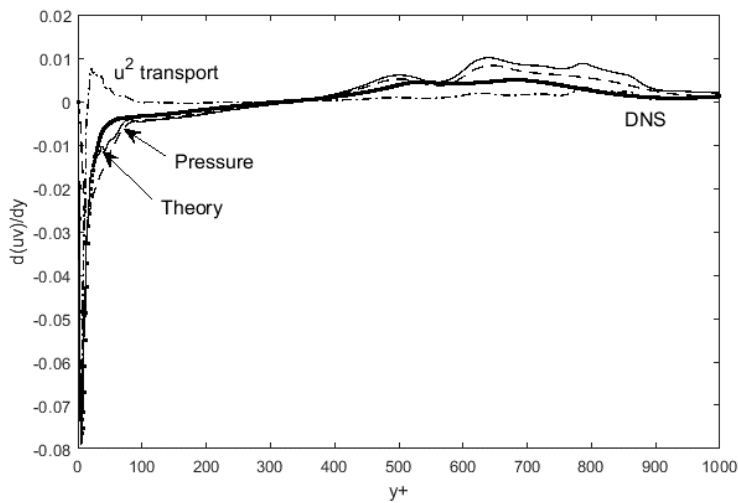

(a)

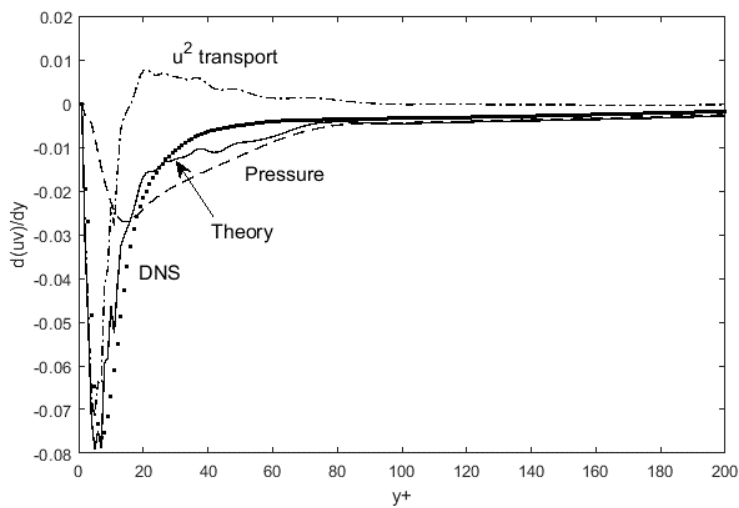

(b)

**Fig. 6. Reynolds stress gradient budget for flow over a flat plate at zero pressure gradient. DNS data is from Kitsios et al. (2016). $Re_{\delta 2} = 3500$. (b) contains the same data as (a), except zoomed on the near-wall region.**

After properly vectorizing the gradients and using the centripetal pressure, we see that the same logic of turbulence momentum transport is applicable in pipe flows with swirl, as shown in Fig. 7. DNS data of Nygard and Andersson (2009) are used, while theoretical (solid) line is from Eq. 13.

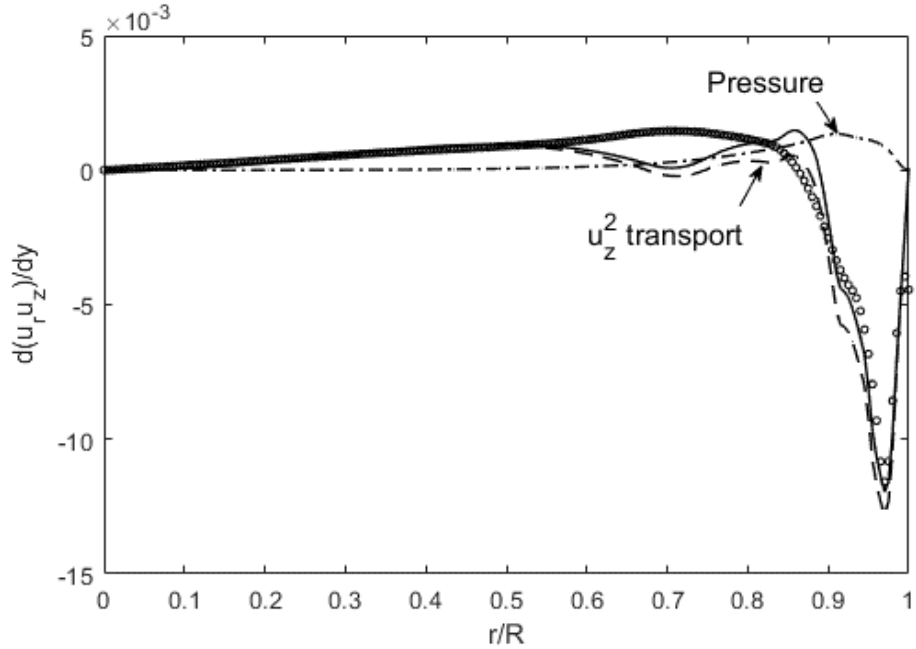

**Fig. 7. Comparison of Reynolds shear stress gradient budget in pipe flows with swirl (Eq. 13) with DNS data (symbols: Nygaard and Andersson, 2009).**

**Use of the Full Turbulent Transport Equations for Jet Flows**

For jet flow with free ambient boundary conditions, the transport equations become yet simpler. Note that the sign convention for jet flows is different from wall-bounded flows.

$$\frac{d(\overline{u'v'})}{dy} = C_1 U \frac{d(\overline{u'^2})}{dy} \tag{15}$$

$$\mu \frac{dU}{dy} = \rho(\overline{u'v'}) \tag{16}$$

$$C_2 U \frac{d(\overline{u'^3})}{dy} = \frac{d(\overline{u'^2 v'})}{dy} + 2v \left(\frac{du'_{rms}}{dy}\right)^2 \qquad (17)$$

This allows for relatively simple numerical solution schemes. Although one was presented in Lee (2020), we describe an improved algorithm to solve the above set of equations. We start by an initial estimate of the u'$^2$ profile and its gradient. Structural constraint method is useful in coming up with this initial estimate, where knowing the peak location and the total energy content as a function of the Reynolds number is quite expedient in structural reconstruction of u'$^2$ profiles (Lee, 2019). With this estimate, initial Reynolds shear stress and the mean velocity profiles can be directly computed using Eqs. 15 and 16, in sequence. The centerline boundary conditions, U/U$_o$= 1, u'$^2$ = (u'$^2$)$_c$, and u'v' =0 at y/y$_H$ = 0, are used, along with the enforcement of the freestream boundary conditions (U → 0, u'v' → 0, and u'$^2$ → 0) to again suppress numerical integration errors. After these steps, u'$^2$ profile can be updated using Eq. 17, and the process iterated until the solution converges, i.e. no further appreciable changes with iterations. The converged results are compared with experimental data (Gutmark and Wygnanski, 1970), in Figs. 8-10.

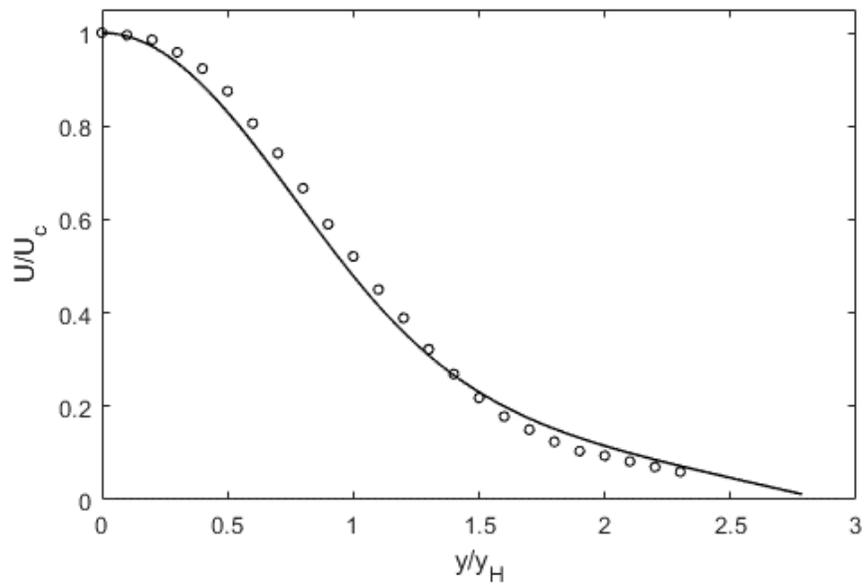

**Fig. 8. Comparison of the mean velocity profile obtained from iterative numerical integration of Eqs. 15-17, with experimental data (Gutmark and Wygnanski, 1970).**

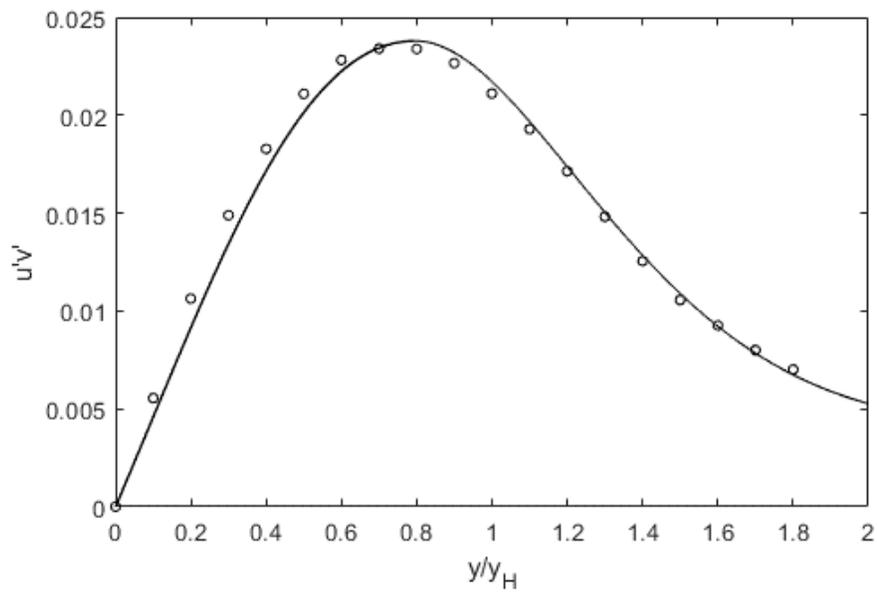

**Fig. 9. Comparison of the Reynolds shear stress obtained from iterative numerical integration of Eqs. 15-17, with experimental data (Gutmark and Wygnanski, 1970).**

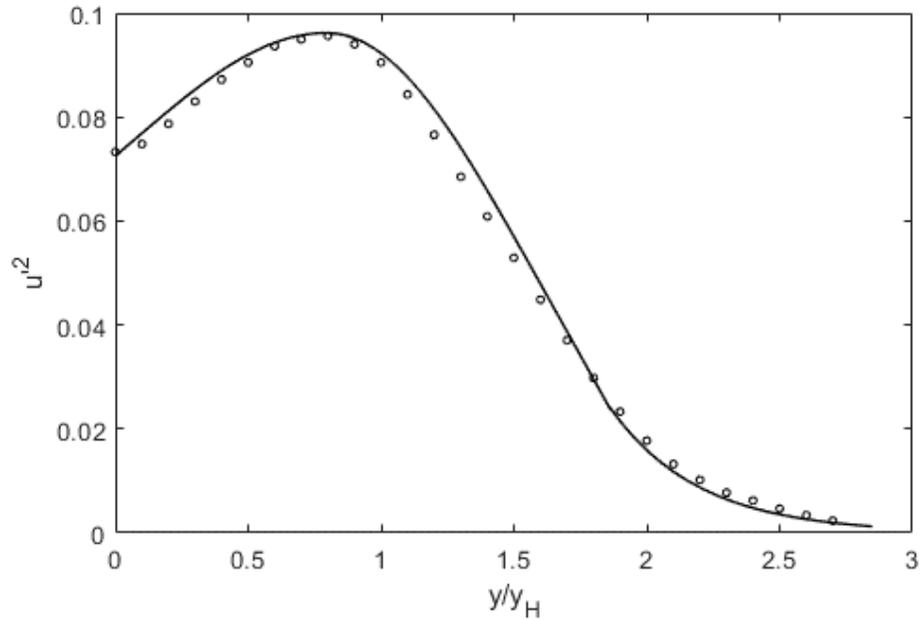

**Fig. 10. Comparison of u'² profile obtained from iterative numerical integration of Eqs. 15-17, with experimental data (Gutmark and Wygnanski, 1970).**

**CONCLUSIONS**

Using the Lagrangian transport of momentum, the Reynolds shear stress can be expressed in terms of basic turbulence parameters. In this view, the Reynolds shear stress emerges as the net sum of the $u'^2$ transport, pressure and shear force terms in the momentum balance. Since the Reynolds stress is a momentum flux term, it is sensible to relate it to other fluctuation momentum terms, as opposed to mean velocity gradients, and this can be expedited in the Lagrangian transport analysis. We extend this formalism to other turbulence parameters such as diagonal components of the Reynolds stress tensor, and also to other geometries (boundary layer flows with adverse pressure gradient and pipe flow with swirl). Data from direct numerical simulations (DNS) are used to validate this full set of turbulent transport equations, exhibiting a good degree of consistency and agreement for all of the components and across different geometries. An example of the use of this full set of turbulence transport equations is shown for turbulent jet flows. This formalism is based on the Lagrangian turbulence transport, and therefore is fundamentally generalizable to other flow configurations. The resulting expressions reproduce the structure of the main turbulence variables, while revealing the internal dynamics linking these components.

**APPENDIX**

**Probe Transform**

For channel flows, the flow is bounded and there is no displacement of the turbulence variables as one travels in the streamwise direction. However, the Galiean transform can be performed at any line of motion, and if we choose a slightly mis-directed path (U* and v*) for the control volume as shown in Figure A1, we obtain the same transform as shown below. In Figure A1, x* and y* axes are aligned in the same direction as U* and v*, respectively.

For a small angle, $\theta \ll 1$, $v^* \ll U$ and $U^* \approx U$. Then,

$$\frac{\partial}{\partial x} = \frac{1}{\cos\theta}\frac{\partial}{\partial x_*} \approx \frac{\partial}{\partial x_*} \tag{A1}$$

$$\frac{\partial}{\partial y} = \frac{1}{\cos\theta}\frac{\partial}{\partial y_*} \approx \frac{\partial}{\partial y_*} \tag{A2}$$

For variable, f, we have

$$\frac{\frac{\partial f}{\partial y_*}}{\frac{\partial f}{\partial x_*}} \approx \frac{\frac{\partial f}{\partial y_*}}{\frac{\partial f}{\partial x_*}} = \tan\theta = \frac{v^*}{U^*} \approx \frac{v^*}{U} \tag{A3}$$

Thus, using this offset transform, we obtain

$$\frac{\partial f}{\partial x} = \frac{U*}{v*}\frac{\partial f}{\partial y} \approx C_1 U \frac{\partial f}{\partial y} \tag{A4}$$

$C_1$ is a constant in the order and unit of v*.

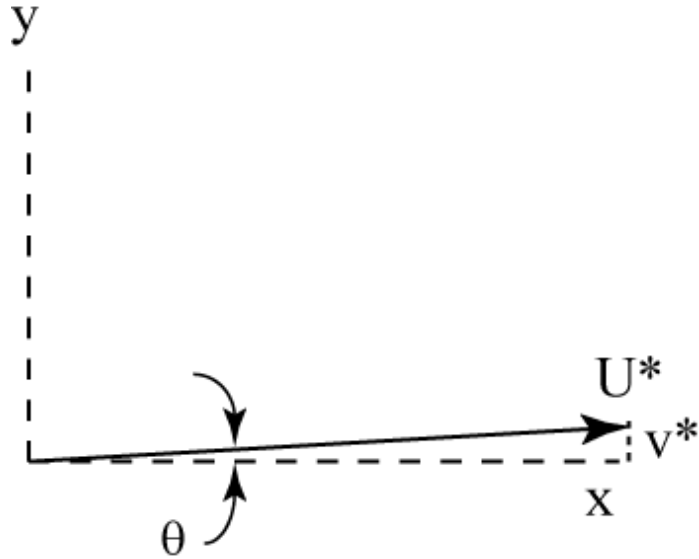

**Figure A1.** Off-set line of motion for the control volume, for "probing" the d/dy gradient.

**REFERENCES**

Graham, J., Kanov, K, Yang, X.I.A., Lee, M.K., Malaya, N, Lalescu, C.C., Burns, R., Eyink, G, Szalay, A, Moser, R.D., and Meneveau, C., A Web Services-accessible database of turbulent channel flow and its use for testing a new integral wall model for LES, Journal of Turbulence, 2016, 17(2), 181-215.

Gutmark, E. and I. J. Wygnanski, The Planar Turbulent Jet, Journal of Fluid Mechanics, 1970, Vol. 73, Part 3, pp. 466-495.

Hamba, F., Exact transport equation for local eddy viscosity in turbulent shear flow, Physics of Fluids, 2013, 25, 085102.

Hanjalic, K., Advanced turbulence closure models: a view of current status and future prospects, Int. Journal of Heat and Fluid Flow, Vol. 15, No. 3, 1994, pp. 178-203.

Inagaki, K., Ariki, T., and Hamba, F., Higher-order realizable algebraic Reynolds stress modeling based on the square root tensor, 2019, arXiv:1905.06106v2.

Kitsios, V, Atkinson, C, Sillero, J.A., Borrell, G, Gungor, A.G., Jiménez, J., and Soria, J, Direct numerical simulation of a self-similar adverse pressure gradient turbulent boundary layer, International Journal of Heat and Fluid Flow, 2016, 61, 129–136.

Launder, B.E., Reece, G.J., and Rodi, W., Progress in the development of a Reynolds-stress closure, Journal of Fluid Mechanics, 1975, Vol. 68, Part 3, pp. 537-566.

Lee, T.-W., Reynolds stress in turbulent flows from a Lagrangian perspective, Journal of Physics Communications, 2018, 2, 055027.

Lee, T.-W., Maximum entropy method for solving the turbulent channel flow problem, 2019, Entropy, 21(7), 675.

Lee, T.-W., Lagrangian Transport Equations and an Iterative Solution Method for Turbulent Jet Flows, Physica D, 2020, 132333.

Lee, T.-W., and Park, J.E., Integral Formula for Determination of the Reynolds Stress in Canonical Flow Geometries, Progress in Turbulence VII (Eds.: Orlu, R, Talamelli, A, Oberlack, M, and Peinke, J.), pp. 147-152, 2017.

Mansour, N.N., Kim, J. and Moin, P., Reynolds-stress and dissipation-rate budgets in a turbulent channel flow, Journal of Fluid Mechanics, 1988, Vol. 194, pp. 15-44.

Nygard, F. and Andersson, H., DNS of swirling turbulent pipe flow, 2009, International Journal of Numerical Methods in Fluids, 64 (9), pp. 945-972.

Pope, S. B., Simple models of turbulent flows, Physics of Fluids, 2011, 23, 011301.

Pope, S. B., Turbulent Flows, Cambridge University Press, 2012.